\documentclass[runningheads]{llncs}
\usepackage[T1]{fontenc}
\usepackage{graphicx}
\usepackage{booktabs}
\usepackage{multirow}
\usepackage[table]{xcolor}
\usepackage{amsmath}
\usepackage{amssymb}
\usepackage{appendix}
\usepackage{float}
\usepackage{subcaption}
\usepackage{hyperref}
\newcommand{\eg}{\textit{e.g.}}

\begin{document}
%
\title{Are Vision Foundation Models Ready for Out-of-the-Box Medical Image Registration?}
%
%
\author{
Hanxue Gu\textsuperscript{1,*} \and
Yaqian Chen\textsuperscript{1,*} \and
Nicholas Konz\textsuperscript{1} \and
Qihang Li\textsuperscript{2} \and
Maciej A. Mazurowski\textsuperscript{3}
}
\index{Gu, Hanxue}
\index{Chen, Yaqian}
\index{Konz, Nicholas}
\index{Li, Qihang}
\index{Mazurowski, Maciej A.}

\institute{
\textsuperscript{1}Department of Electrical and Computer Engineering, Duke University \\
\textsuperscript{2}Department of Biostatistics and Bioinformatics, Duke University \\
\textsuperscript{3}Departments of Biostatistics and Bioinformatics, Radiology, Electrical and Computer Engineering, and Computer Science, Duke University \\
\textsuperscript{*}Equal contribution
}
\maketitle              
\begin{abstract}
Vision foundation models (VFM), pre-trained on large image datasets and capable of capturing rich feature representations, have recently shown potential for zero-shot image registration. However, their performance has mostly been tested in the context of less complex structures, such as the brain or abdominal organs, and it remains unclear whether these models can handle more challenging, deformable anatomy. Breast MRI registration is particularly challenging due to anatomical variation between patients, deformation from positioning, and the thin, complex structure of fibroglandular tissue (FGT), where accurate alignment is crucial.
In this study, we provide a comprehensive evaluation of VFM-based registration algorithms. 
We assess five pre-trained encoders, including DINO-v2, SAM, MedSAM, SSLSAM, and MedCLIP, across \textbf{four} key breast registration tasks that capture variations in different years and dates, sequences, modalities, and patient disease status (lesion versus no lesion). Our results show that among the five pre-trained encoders, SAM demonstrates a clear advantage. VFM outperforms traditional optimization-based methods, especially in cross-modality tasks, with its advantage more pronounced for large structures (e.g., organs and breast) than for finer anatomical details like FGT. Further work is needed to understand how domain-specific training influences registration and to explore targeted strategies that improve fine structure accuracy.
We also publicly release our code at \href{https://github.com/mazurowski-lab/Foundation-based-reg}{Github}.


\end{abstract}
\section{Introduction}
Accurate alignment of breast MRI images across longitudinal studies, sequences, or even modalities (\eg, PET-CT) is essential for tracking tumor changes and assisting surgery planning \cite{alfano2022breast}. However, deformable registration of the breast remains highly challenging due to the natural variability and high deformability of both the breast and dense tissue \cite{mehrabian2018deformable,chen2025guidedmorph}. Most studies still rely on traditional optimization-based registration algorithms \cite{ringel2022supine}, such as the symmetric image normalization method (SyN) \cite{avants2008symmetric} and Elastix \cite{klein2009elastix}. Several deep learning-based registration methods have recently been proposed, but they typically require large, task-specific training datasets or annotated masks, making them less practical or transferable across different clinical settings \cite{balakrishnan2019voxelmorph,chen2022transmorph,kim2021cyclemorph}.
\begin{figure}
    \centering
    \includegraphics[width=0.8\linewidth]{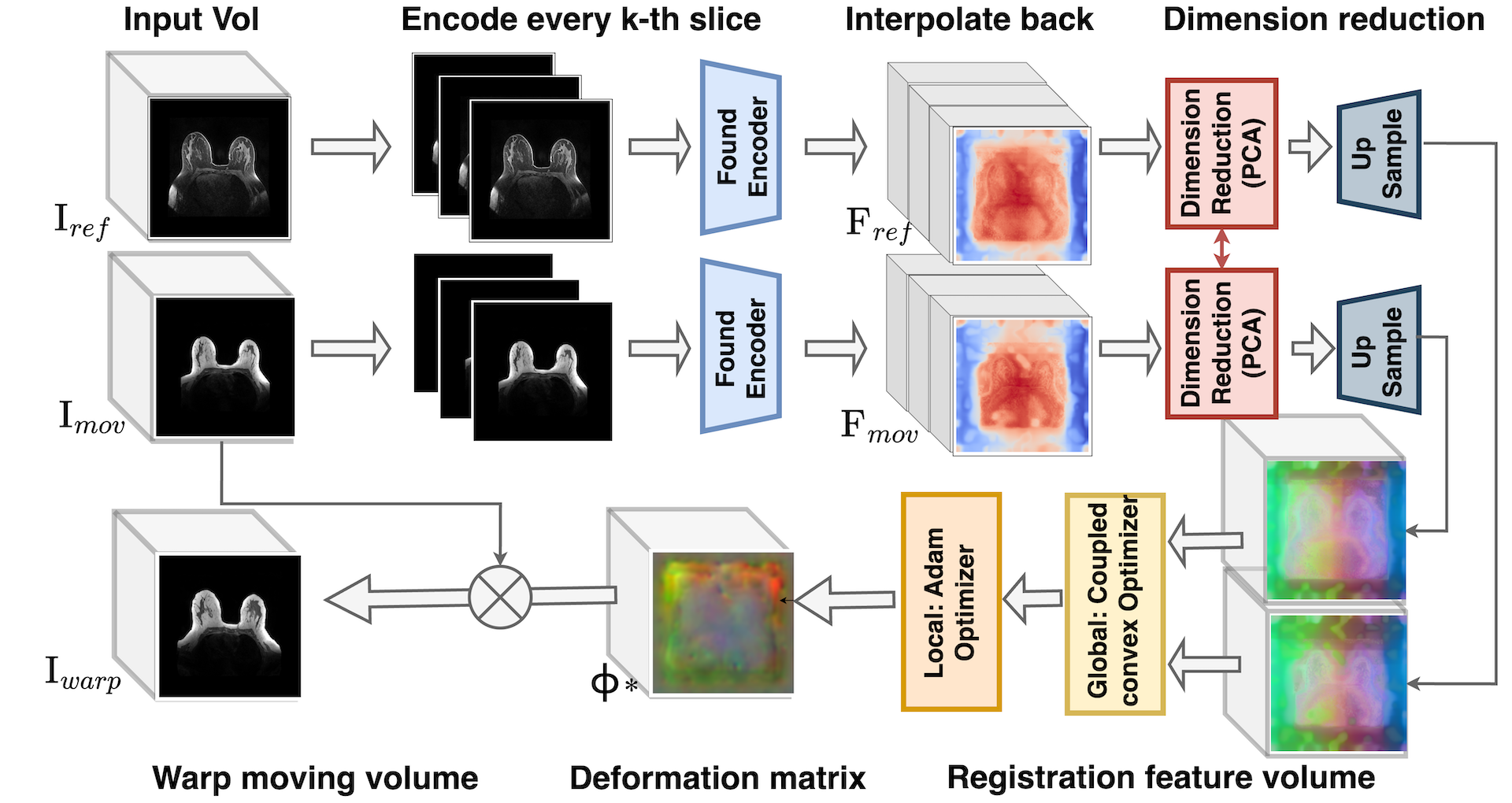}
    \caption{Pipeline for VFM-based Breast Image Registration. The Found Encoder can be replaced with any pretrained vision foundation encoder.}
    \label{fig:pipeline}
\end{figure}

In contrast, vision foundation models (VFMs) have gained attention for their strong generalization ability and zero-shot performance on downstream tasks, such as segmentation and classification \cite{shi2023generalist,mazurowski2023segment,dong2024segment}. One recent effort, DINO-Reg \cite{song2024dino}, shows that DINO-v2’s rich semantic embeddings can support effective zero-shot registration across medical imaging domains. Yet, it remains unclear how different pre-training strategies, such as masked autoencoders (MAE), vision-language models (VLMs), or medical domain-specific models like MedSAM, affect registration performance. Moreover, most studies focus on less deformable regions like abdominal organs with larger, more stable structures, while breast registration poses unique challenges with significant shape variation and thin, highly deformable tissue. How VFM-based algorithms handle such complex anatomy remains poorly understood.

To mitigate this gap, we present the first systematic benchmark of various foundation model encoders for zero-shot breast MRI registration. Our study evaluates different types of pretraining strategies, compares foundation-based methods with other zero-shot approaches, and examines the impact of domain-specific versus natural image pretraining. We conduct comprehensive experiments on a curated breast MRI dataset across multiple challenging scenarios, including multi-time same-sequence registration, multi-sequence registration, cross-modality alignment, and lesion tracking. 

The main contributions of our work are:
\begin{enumerate}
\item We implement a flexible pipeline for foundation-model-based algorithms supporting different pretrained encoders, and make the code publicly available.
\item We conduct a comprehensive comparison of registration strategies according to the pretraining strategies that were applied to develop the foundation model, including contrastive learning, masked autoencoder, and vision-language approaches, across both natural and medical imaging domains.
\item We design four registration tasks that reflect real-world challenges, covering longitudinal, cross-sequence, lesion-tracking, and PET-to-MRI alignment.
\item We assess model performance across targets of varying size, anatomical detail, and deformation, including breast, dense tissue, organ, and lesion.
\item We systematically evaluate the strengths and limitations of VFM-based algorithms compared to other zero-shot approaches.
\end{enumerate}

\section{Method}
We adopt a training-free registration pipeline that extracts semantic embeddings from input volumes using a pretrained foundation model encoder, followed by deformable registration performed on the embeddings (dimension-reduced) without any additional training or fine-tuning (Sec. ~\ref{sec:pipeline}). The pre-trained encoders can be flexibly used as plug-in components. These encoders vary in pre-training strategies (e.g., contrastive learning or masked autoencoding) and in whether they were pre-trained on medical images or natural images (Sec. ~\ref{sec:encoder-choice}).
\subsection{VFM-based registration}\label{sec:pipeline}
Our training-free, VFM-based registration pipeline, inspired by the framework of ~\cite{song2024dino}, includes three main stages: (1) feature extraction, (2) dimension reduction, and (3) feature-based registration, as shown in Fig. \ref{fig:pipeline}.

\noindent\textbf{Feature extraction}. Let $I_{\text{ref}}$ and $I_{\text{mov}}$ denote reference and moving image volumes for registration. Considering that most available pre-trained encoders are for 2D images, we perform feature extraction slice-by-slice along the axial plane. Each volume is treated as a sequence of 2D axial slices, denoted $\{I_{\text{ref}}^z\}_{z=1}^{Z}$ and $\{I_{\text{mov}}^z\}_{z=1}^{Z}$, where $z$ indexes the slice number.
To reduce computational cost, we chose to encode every $k$-th slice, where $k$ is a hyperparameter that can be selected based on the desired speed of registration. The selected slices are passed through a frozen vision encoder $E(\cdot)$, producing intermediate feature maps:
\begin{equation}
  F_{\text{ref}}^z = E(I_{\text{ref}}^z), \quad F_{\text{mov}}^z = E(I_{\text{mov}}^z), \quad \text{for } z \in \{1, k+1, 2k+1, \ldots\}.
\end{equation}

The features for the skipped slices are then estimated via linear interpolation along the slice axis to form a complete 3D feature volume. When $k=1$, all slices are encoded and no interpolation is applied. Here, $E(\cdot)$ denotes the encoder part of foundation models (e.g., DINO-v2, SAM), typically based on vision transformers (ViTs). For each encoder, the output feature map $F^z \in \mathbb{R}^{N \times D}$ consists of $N$ patch tokens and, optionally, a global CLS token, each of dimension $D$.

\noindent\textbf{Dimension reduction}. In the latent space, the feature dimension $D$ is typically high (e.g., 256 for SAM and 384 for DINO-v2), which can introduce noise and lead to significant computational cost during registration. To address this, we apply principal component analysis (PCA) to reduce the feature dimensionality for each patch token.
Let $F^z \in \mathbb{R}^{N \times D}$ denote the feature matrix of slice $z$, where $N$ is the number of patch tokens. We collect features from all sampled slices of both the reference and moving volumes and concatenate them along the patch dimension, resulting in a joint matrix  $F_{\text{all}} \in \mathbb{R}^{M \times D}$,  where $M = N \times Z \times 2$.
PCA is then computed on this combined matrix, and the resulting projection is applied slice-wise to both volumes. Let $d$ be the number of retained principal components (e.g., $d = 12$), the reduced feature maps become:
\begin{equation}
\tilde{F}_{\text{ref}}^z = F_{\text{ref}}^z \cdot W_{\text{PCA}}, \quad \tilde{F}_{\text{mov}}^z = F_{\text{mov}}^z \cdot W_{\text{PCA}}, \quad \tilde{F}_{\text{ref}}^z, \tilde{F}_{\text{mov}}^z \in \mathbb{R}^{N \times d},
\end{equation}
Here, $W_{\text{PCA}} \in \mathbb{R}^{D \times d}$ is the projection matrix computed from PCA on $F_{\text{all}}$. This ensures both volumes are embedded into the same reduced feature space.

To prepare for volumetric registration, we reshape $\tilde{F}_{\text{ref}}^z$, $\tilde{F}_{\text{mov}}^z$ back to a 3D spatial format. Specifically, each slice-level feature map $\tilde{F}^z \in \mathbb{R}^{N \times d}$ is first reshaped into a 2D grid of size $w \times h \times d$, where $w$ and $h$ represent the number of patch tokens along the width and height, respectively. 
Stacking the reshaped slices along the $z$-axis results in a 4D volume. This intermediate volume is then upsampled to match the original spatial resolution, resulting in a final feature volume of size $W \times H \times Z \times d$, which aligns with the input dimensions and serves as input to the registration step.

\noindent\textbf{Feature-based Registration.} We follow ConvexAdam~\cite{siebert2024convexadam}, adopting a two-stage registration strategy that combines global discrete optimization with instance-level refinement to efficiently align the reduced feature volumes.

In the first stage, we employ a coupled convex discrete optimization framework to obtain an initial deformation field based on the extracted feature volumes. A convex energy function is formulated, which combines the feature similarity term with a regularization term to ensure smooth deformations:
\begin{equation}
  \mathcal{E}(\phi) = - \text{Sim}\left( \tilde{F}_{\text{ref}}, \tilde{F}_{\text{mov}} \circ \phi \right) + \lambda \|\nabla \phi\|^2  
\end{equation}
Here, $\phi$ is the dense displacement field, $\text{Sim}(\cdot,\cdot)$ denotes a similarity metric (e.g., sum of squared differences or local cross-correlation), and $\lambda$ controls the smoothness constraint. The convex optimization is performed within a discrete search space to efficiently identify the globally optimal deformation. In this work, we set $\lambda = 2$ and MSE for $Sim$, consistent with DINO-Reg \cite{song2024dino}.

In the second stage, the deformation field is refined through continuous instance-wise optimization via Adam \cite{adam}. Initialized with the displacement field from the first stage, we iteratively update $\phi$ by minimizing the same energy function in a differentiable manner, allowing for sub-voxel accuracy and capturing fine-scale deformations: $\phi^{*} = \underset{\phi}{\arg\min} \; \mathcal{E}(\phi)$.
This optimization strategy utilizes the robustness of discrete search for global alignment and the flexibility of continuous optimization for local refinement.

\subsection{Foundation models}\label{sec:encoder-choice}
\begin{figure}[h]
    \centering
    \includegraphics[width=0.9\linewidth]{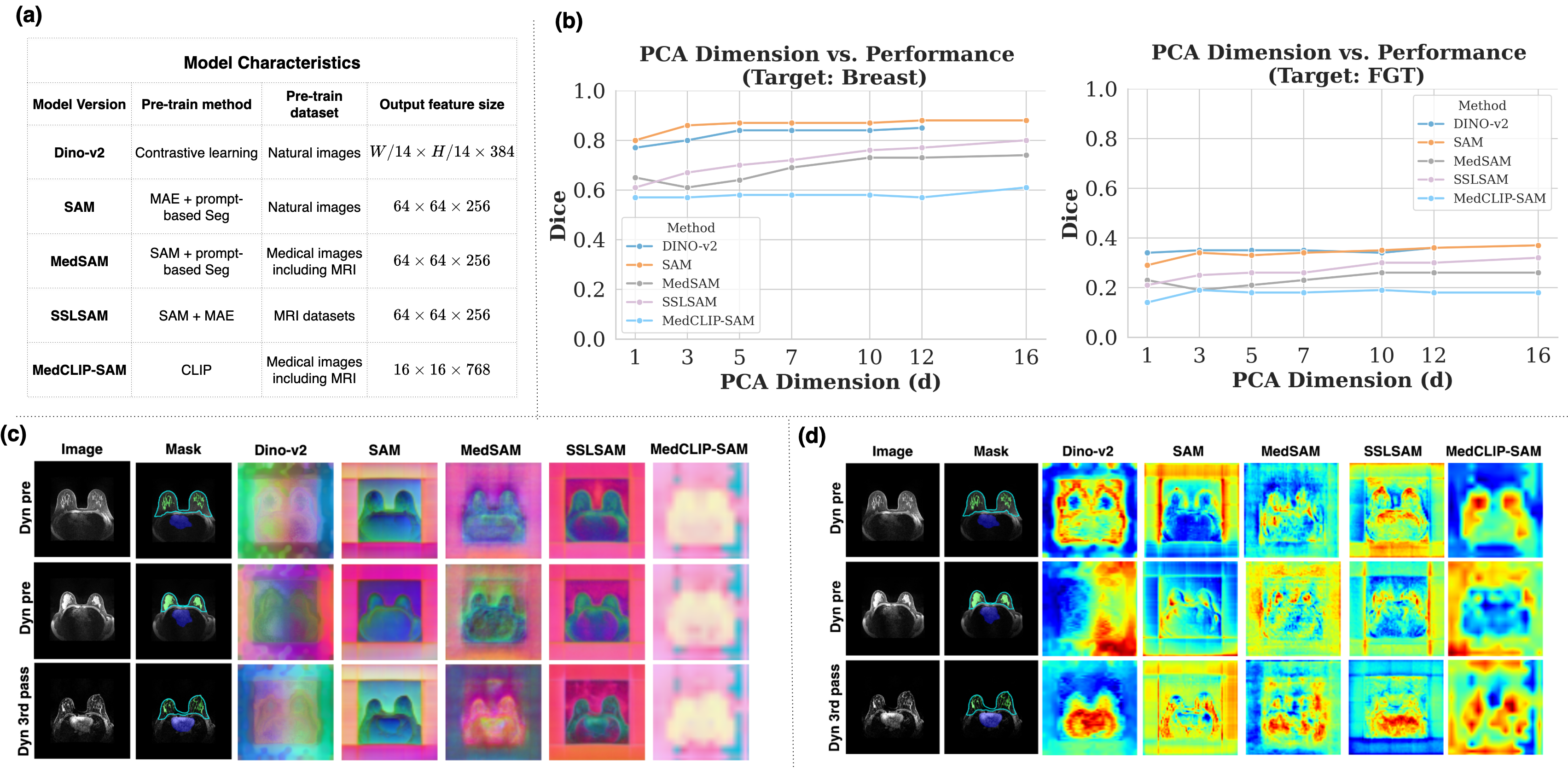}
    \caption{(a) Encoder characteristics for each VFM (training strategy and dataset domain). (b) Ablation of dimensionality reduction (PCA dimension $d$) versus registration performance for each encoder. (c) First three channels of the dimension-reduced feature map, visualized as RGB. (d) Top feature channel from raw feature map showing the highest FGT contrast, aiming to visualize whether FGT-related information is captured in the latent features.}
    \label{fig:feature}
\end{figure}

We use five publicly available foundation models as feature extractors, varying in pre-training strategy and domain (natural vs. medical), as shown in Fig. \ref{fig:feature}.

\noindent\textbf{DINO-v2 \cite{oquab2024dinov2learningrobustvisual}}: Pre-trained on natural images using a self-supervised contrastive learning strategy (self-distillation without labels). DINO-v2 is known for producing semantically rich features and has been applied in one prior work for cross-domain medical image registration~\cite{song2024dino}.

\noindent\textbf{SAM \cite{kirillov2023segment}}: Trained on large-scale natural image segmentation datasets using a promptable segmentation objective. The model adopts a ViT-based encoder and was first pre-trained with MAE, followed by additional prompt-based segmentation training. SAM has also been widely used for general vision tasks. Although SAM is designed for prompt-based segmentation, we use only its image encoder to automatically generate latent embeddings without human input.

\noindent\textbf{MedSAM \cite{ma2024segment}}: A domain-specific version of SAM, fine-tuned on various medical image datasets including MRIs under a prompt-based setting to improve performance on medical image segmentation tasks. Notably, as shown in the original MedSAM Supplementary Tables 1–4, the training datasets did not include \textit{any} breast MRI or breast CT images.

\noindent\textbf{SSLSAM \cite{gu2024build}}: Fine-tuned from SAM's initial weights on various MRI datasets without labels under the MAE setting. Noticeably, their pre-trained dataset contains one publicly available breast dataset for dynamic pre-contrast MRIs, but not other breast MRI sequences, \eg, post-contrast \cite{lew2024publicly}.

\noindent\textbf{MedCLIP-SAM \cite{koleilat2025medclipsamv2universaltextdrivenmedical}}: A vision-language foundation model trained on paired medical images and radiology reports. It follows the same training strategy as MedCLIP, but while MedCLIP was only trained on radiographs, MedCLIP-SAM includes additional MRI data during pretraining. 


\section{Experiments}
\subsection{Datasets preparation}
\noindent\textbf{Exam collection.}\label{sec:appendix:exam}
For the MRI-to-MRI registration tasks, the data is collected internally from our institution, consisting of exams performed under the bilateral breast MRI protocol, ranging from the year 2014 to the year 2021. We only include patients who have multiple time-stamped exams and at least one pre-contrast axial view volume available. 
For the PET-CT to MRI registration task, the data is collected externally from BREAST-DIAGNOSIS collection \cite{bloch2015breastdiagnosis}, where we select patients who have both PET-CT and MRI scans acquired within a six-month interval.
All exams are first resampled into $1mm \times 1mm \times 1mm$ isotropic voxel spacing and then padded into $512 \times 512 \times 512$ as volume size. For each task, we randomly selected fifty pairs from the pairs that meet the task criteria, with the task details described in a later paragraph \textit{Tasks}. From these pairs, volumes from patients who had undergone mastectomy were excluded. 

\noindent\textbf{Mask collection.}\label{sec:appendix:mask}
We include three targets for our registration evaluation: breast, FGT, and organ. The first two, breast and FGT, are important targets to register for breast-specific tasks, as accurate alignment of both overall breast shape and internal dense tissue is critical for clinical and research use. For the organ label, we suspect that pre-trained networks may perform relatively well, as this type of structure is commonly seen in abdominal MRI or CT datasets that many of these models were originally exposed to. 
For breast and FGT mask, we employ the breast segmentation model given by the previous study utilizing 3D V-Net architecture \cite{chen2025breast,lew2024publicly} followed by the manual correction. 
The organ and lesion masks are generated using a publicly available breast MRI segmentation model based on nnU-Net, followed by manual correction. The generated masks are visualized in the right-hand side of Fig. \ref{fig:feature}, in the \textit{Mask} column.

\noindent\textbf{Tasks design.}
In this study, we design \textit{four} tasks and one ablation study to comprehensively evaluate the registration performance of multiple models. \textit{Task 1} involves registering breast MRI scans from different dates and years (longitudinal exams) with the same image sequence (all images are dynamic pre-contrast). \textit{Task 2} focuses on registering longitudinal breast MRI exams with different image sequences, where the fixed images are set to the dynamic pre-contrast sequence. The category of moving image sequences is shown in Fig. \ref{fig:vis-result} (b). \textit{Task 3} explores lesion tracking, where we register the moving image with the lesion to the fixed image without the lesion. This task is designed to evaluate whether the registration model attempts to erase or suppress the lesion in the moving image in order to match the target. For \textit{Task 4}, we register PET-CT to MRI, which is more challenging due to both the cross-modality nature and the significant breast deformation caused by different patient postures (lying on the back versus facing down). We also include the ablation study to disentangle the effects of feature representation and optimization strategy, we replaced ConvexAdam with standard registration methods, SyN and NiftyReg FFD, to register SAM-derived feature volumes in three randomly selected volume pairs from Task 2. The result for all tasks are shown in Table \ref{tab:quant_result}.

\noindent\textbf{Evaluation metrics.}
\textit{Tasks 1} to \textit{Task 3} are evaluated using the Dice Similarity Coefficient ($DSC$) of three targets: breast, FGT, and organ. 
For \textit{Task 3}, we further assess lesion preservation by calculating $V_{\varepsilon} = abs(\sum M_{les-warped} - \sum M_{les-mov}) / \sum M_{les-mov}$, which reflects the change in lesion size during registration. For lesion tracking purposes, the ideal goal is to align the breast and dense tissue while preserving the lesion’s physical characteristics, rather than altering its shape or size during the registration process. For \textit{Task 4}, we qualitatively evaluate the registration performance by visualizing the warped CT and overlaying the PET onto the fixed MRI.

\subsection{Implementation details of VFM-based registration}
For SAM-related models, including SAM, MedSAM, and SSL-SAM, the input slices are resized to $1024 \times 1024$. For DINO-v2, the input size is flexible, so we upsample the slices to the largest size that can fit into a single GPU A6000, which is $1792 \times 1792$. For MedCLIP-SAM, the input slices are resized to $224 \times 224$. All inputs are normalized according to the pretraining settings of each model, using the normalization methods provided in their respective codebases.
For the \textit{dimension reduction} step, the size of the extracted feature map for each slice ($F^z$) varies by model. 
The PCA dimension $d$ for each encoder is optimized on 9 sampled pairs from \textit{Task 3}. 
We select the $d$ at which DSC plateaus, balancing performance and computational cost, as higher $d$ increases memory usage and slows registration (Fig. \ref{fig:feature}).
Based on this, we set $d=16$ for SAM, $d=10$ for MedSAM, $d=16$ for SSLSAM, $d=10$ for DINO-v2, and $d=16$ for MedCLIP-SAM.
The choice of $d$ is based on an ablation from $d=1$ to $16$ (capped at $d=12$ for DINO-v2 due to GPU limits).
For the \textit{registration} step, we follow the hyperparameters selected by DINO-reg \cite{song2024dino}.






\section{Results and Discussion}
\renewcommand{\arraystretch}{0.6} 
\begin{table}[h]
\setlength{\tabcolsep}{4pt}
\centering
\fontsize{8pt}{8pt}\selectfont
\caption{Comparison of Methods on Three Targets Using DSC(\%) and Error Size(\%).}
\label{tab:quant_result}
\begin{tabular}{ll|c|c|c|c}
\toprule
\multicolumn{2}{c|}{\multirow{2}{*}{\textbf{Method}}} & \multicolumn{1}{c|}{\textbf{Breast}} & \multicolumn{1}{c|}{\textbf{FGT}} & \multicolumn{1}{c|}{\textbf{Organ}} & \textbf{Lesion}\\
\cline{3-6}
\multicolumn{2}{c|}{} & DSC $\uparrow$ &  DSC $\uparrow$ & DSC $\uparrow$ & $V_{\varepsilon}$ $\downarrow$\\
\midrule
\rowcolor{gray!20}
\multicolumn{6}{l}{\textbf{Task 1: Registration between longitudinal scans within the same MRI sequence}} \\
\multirow{1}{*}{\textbf{Init.}} & Raw & 49.0 $\pm$ 20.6 & 21.2 $\pm$ 16.6 & 48.3 $\pm$ 21.5 & - \\\midrule
\multirow{3}{*}{\textbf{Opt.}} & Affine & 68.2 $\pm$ 21.7 & 33.0 $\pm$ 21.3 & 60.9 $\pm$ 22.8 & -  \\
                                & Syn & 80.0 $\pm$ 20.0 & 45.4 $\pm$ 24.9 & 66.7 $\pm$ 23.6 & -  \\
                                & NiftyReg & 81.6 $\pm$ 17.9 & 53.8 $\pm$ 26.5 & 69.2 $\pm$ 24.2 & -  \\
                                & ConvexAdam & 80.8 $\pm$ 16.6 & 53.3 $\pm$ 23.4 & 64.7 $\pm$ 21.3 & -\\
\midrule
\multirow{4}{*}{\textbf{Found.}}  & DINO-v2  & 81.3 $\pm$ 11.7 & 47.4 $\pm$ 21.8 & 66.6 $\pm$ 19.4 & -  \\
                                  & SAM  & \textbf{87.0 $\pm$ 8.3} & 53.8 $\pm$ 17.6 & 69.1 $\pm$ 18.4 & -  \\
                                  & MedSAM   & 80.9 $\pm$ 11.3 &  47.3 $\pm$ 21.6 & 68.9 $\pm$ 18.3 & -  \\
                                  & SSLSAM  &  82.1 $\pm$ 13.7 & \textbf{55.0 $\pm$ 22.7} & \textbf{72.3 $\pm$ 16.9} & -  \\
                                  & MedCLIP-SAM  & 69.7 $\pm$ 13.8 & 33.4 $\pm$ 21.5 & 59.1 $\pm$ 16.1 & -  \\
\midrule
\rowcolor{gray!20}
\multicolumn{6}{l}{\textbf{Task 2: Registration between longitudinal scans with different MRI sequences}} \\
\multirow{1}{*}{\textbf{Init.}} & Raw & 44.1 $\pm$ 24.4 & 18.7 $\pm$ 13.4 & 45.3 $\pm$ 21.6 &  - \\
\midrule
\multirow{3}{*}{\textbf{Opt.}} & Affine & 65.8 $\pm$ 22.3 & 32.0 $\pm$ 18.5 & 61.1 $\pm$ 22.7 & - \\
                               & Syn    & 76.8 $\pm$ 19.1 & 42.8 $\pm$ 21.7 & 64.4 $\pm$ 21.6 & - \\
                               & NiftyReg    & 73.9 $\pm$ 22.2 & 44.3 $\pm$ 25.0 & 62.1 $\pm$ 22.0 & - \\
                               & ConvexAdam  & 73.4 $\pm$ 10.3 & 42.4 $\pm$ 31.9 & 41.2 $\pm$ 35.4 & -\\
                                      
\midrule
\multirow{4}{*}{\textbf{Found.}}  & DINO-v2  &  80.4 $\pm$ 11.4 & 44.5 $\pm$ 18.9 & 62.2 $\pm$ 18.9 & - \\
                                  & SAM      & \textbf{85.3 $\pm$ 9.0} & \textbf{50.2 $\pm$ 19.4} & 66.4 $\pm$ 17.7 & - \\
                                  & MedSAM   & 76.3 $\pm$ 15.1 & 40.0 $\pm$ 19.0 & \textbf{66.5 $\pm$ 18.3} & - \\
                                  & SSLSAM   & 81.5 $\pm$ 12.1 & 45.7 $\pm$ 19.1 & 63.6 $\pm$ 21.6 & - \\
                                  & MedCLIP-SAM  & 59.8 $\pm$ 18.7 & 18.5 $\pm$ 15.9 & 50.6 $\pm$ 16.3 & - \\
\midrule
\rowcolor{gray!20}
\multicolumn{6}{l}{\textbf{Task 3: Registration from lesion-present cases to lesion-absent cases}} \\
\multirow{1}{*}{\textbf{Init.}} & Raw & 49.4 $\pm$ 20.3 & 18.1 $\pm$ 14.3 & 40.4 $\pm$ 21.9 & 0 (ref) \\
\midrule
\multirow{3}{*}{\textbf{Opt.}} & Affine & 68.0 $\pm$ 15.6 & 31.2 $\pm$ 18.2 & 53.8 $\pm$ 24.4 & 0 (ref) \\
                               & Syn   & 80.5 $\pm$ 10.1 & 41.5 $\pm$ 21.5 & 57.6 $\pm$ 24.8 & 37.7 $\pm$ 28.6\\
                               & NiftyReg & 81.0 $\pm$ 9.9 & 48.4 $\pm$ 22.3 & 57.5 $\pm$ 25.5 & 48.9 $\pm$ 30.3\\
                               & ConvexAdam & 78.3 $\pm$ 18.9 & \textbf{50.3 $\pm$ 24.4} & 53.3 $\pm$ 27.3 & 60.5 $\pm$ 85.5\\
\midrule
\multirow{4}{*}{\textbf{Found.}}  & DINO-v2  & 80.5 $\pm$ 10.6 & 40.8 $\pm$ 18.7 & 55.1 $\pm$ 22.0 & \textbf{31.3 $\pm$ 23.3}\\
                                  & SAM      & \textbf{86.9 $\pm$ 8.6} & 46.7 $\pm$ 21.2 & \textbf{60.0 $\pm$ 23.9} & 48.4 $\pm$ 32.8\\
                                  & MedSAM   & 79.7 $\pm$ 12.6 & 39.3 $\pm$ 21.0 & 59.3 $\pm$ 23.7 & 46.7 $\pm$ 32.2\\
                                  & SSLSAM   & 85.2 $\pm$ 9.9 & 45.7 $\pm$ 21.2 & 59.1 $\pm$ 23.7 & 60.9 $\pm$ 41.2\\
                                  & MedCLIP-SAM  & 69.6 $\pm$ 17.9 & 26.6 $\pm$ 18.9 & 49.0 $\pm$ 21.4 & 98.6 $\pm$ 150.8 \\     
\midrule
\rowcolor{gray!20}
\multicolumn{6}{l}{\textbf{Ablation study: contribution of features vs. optimizer}} 
\\
& Raw + SyN        & 71.1 $\pm$ 30.4 & 40.9 $\pm$ 18.5 & 65.7 $\pm$ 30.0 & -\\
& Raw + NiftyReg   & 70.5 $\pm$ 20.4 & \textbf{50.5 $\pm$ 19.0} & 70.7 $\pm$ 6.5 & -\\
& Raw + ConvexAdam   & 69.4 $\pm$ 25.4 & 46.2 $\pm$ 22.7 & 63.7 $\pm$ 11.3 & -\\
& SAM + SyN        & 58.9 $\pm$ 32.7 & 24.7 $\pm$ 11.1 & 53.7 $\pm$ 24.8 & -\\
& SAM + NiftyReg   & 74.9 $\pm$ 26.5 & 35.7 $\pm$ 20.7 & 69.7 $\pm$ 10.7 & -\\
& SAM + ConvexAdam & \textbf{85.0 $\pm$ 18.8} & 47.0 $\pm$ 27.9  & \textbf{73.4 $\pm$ 9.2} & -\\
\hline
\end{tabular}
\end{table}
\noindent\textbf{Results.} 
Table~\ref{tab:quant_result} shows that for same-sequence MRI registration (Task 1), foundation models, especially SAM, achieve superior performance in large-structure registration like the breast contour, reaching a DSC of $87.0\%$, and clearly outperforming the best optimization-based method.
For FGT, we observe that SAM and SSLSAM perform better in \textit{Task 1}, while other VFM-based models show a decreased performance.
For organ registration, as we initially hypothesized, models pre-trained on medical-domain data perform better, with SSLSAM achieving the highest performance due to its exposure to similar anatomical structures.

\begin{figure}[h]
    \centering
    \includegraphics[width=0.99\linewidth]{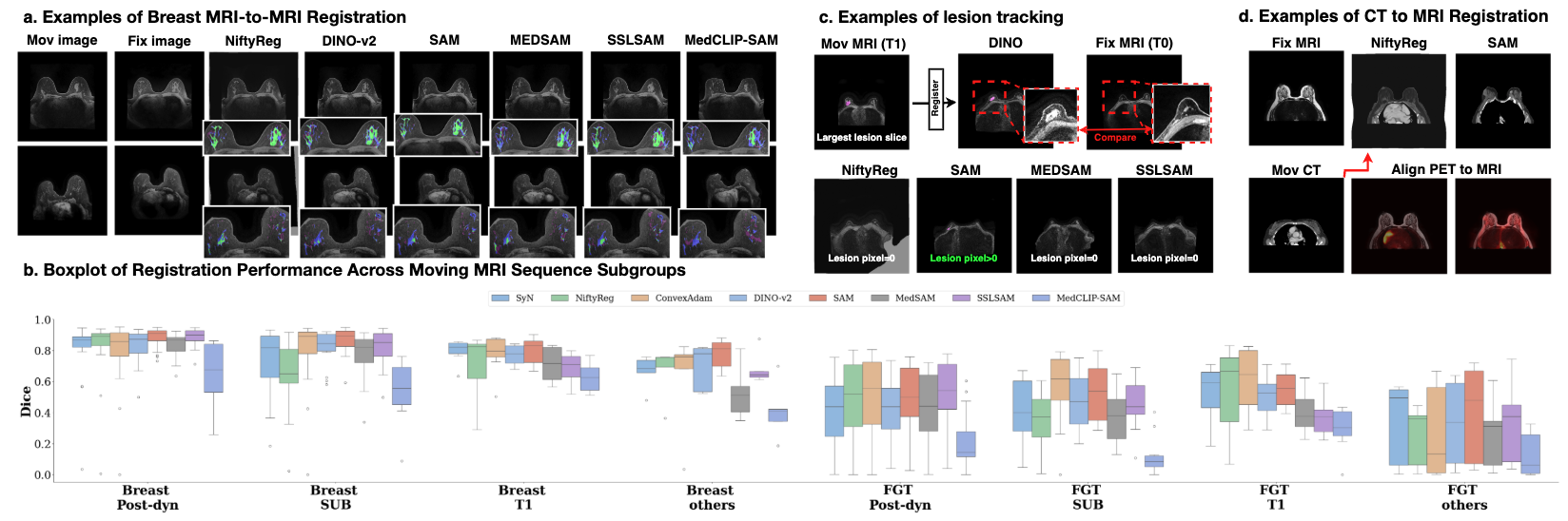}
    \caption{(a). Visualization of example results for different VFM-based algorithms. FGT from the fixed image slice is shown in blue, and the warped dense tissue is shown in purple. The overlapping regions appear in green; (b). subgroup analysis of the registration performance versus the sequence of the moving image; (c)-(d). Visualization results of selected cases for task 3 and task 4. }
    \label{fig:vis-result}
\end{figure}

In cross-sequence registration (\textit{Task 2}), SAM achieves a DSC improvement of $8.5\%$ for breast, $5.9\%$ for FGT, and $2\%$ for organ compared to the best optimization-based method, with further visualizations of performance across different sequence subgroups shown in Fig.~\ref{fig:vis-result} (b).
This suggests that features extracted by foundation models are domain-invariant, making them more robust under shifts in image appearance. In contrast, optimization-based methods rely more heavily on image-level similarity.

Regarding \textit{Task 3} (lesion tracking), DINO-v2 achieves the best performance in preserving lesion size, while MedCLIP-SAM performs poorly. For \textit{Task 4} (PET-CT to MRI registration), the benefits of foundation models become more evident. As shown in Fig. \ref{fig:vis-result} (d), NiftyReg completely fails to align organs between CT and MRI, SAM successfully registers CT to MRI despite significant shape differences, confirming the advantage of foundation models under large domain gaps (e.g., CT to MRI) \cite{song2024dino}. This result also suggests potential applications in overlaying PET onto MRI to visualize tumors.

\noindent\textbf{Discussion for pre-train strategies.} 
Although both SAM and DINO-v2 are pre-trained on natural images, we observe a performance gap here. We attribute this to differences in their training strategies: SAM uses a two-stage pipeline, where the second stage is explicitly trained for pixel-wise segmentation, enabling it to learn fine-grained, spatially localized, and structure-aware representations.
DINO-v2, on the other hand, is trained via contrastive learning. While it includes local and global tokens, the patch sizes are relatively large, limiting its ability to capture detailed spatial information.
Compared to MedCLIP-SAM, the feature tokens are even coarser, and the model is originally designed for image-text alignment rather than local-information preserving. 
To further understand these differences, we visualize the feature maps from each model in Fig.\ref{fig:feature}. For row 3, only SAM and SSLSAM capture some FGT structures, while others fail. 
Even for SAM/SSLSAM, FGT details is barely visible in the feature map, this limitation is also evident in the qualitative results (Fig.~\ref{fig:vis-result} (a)), where models generally register the breast contour well, but substantial FGT misalignment remains. The zero-shot performance of the best VFM-based method still lags significantly behind other training-based approaches, such as GuidedMorph~\cite{chen2025guidedmorph}.

\noindent\textbf{Discussion for pre-train datasets.} 
When comparing models trained on natural images versus those trained on medical datasets (e.g., MedSAM) or pre-contrast breast MRIs (SSLSAM), we observe mixed results.
SSLSAM shows improvement in FGT and organ registration in \textit{Task 1}, but performance drops in \textit{Task 2}. We suspect this is due to SSLSAM’s pretraining solely on pre-contrast breast MRIs \cite{gu2024build}, leading to domain overfitting. As a result, while SSLSAM performs reasonably well for pre-contrast MRI alignment (\textit{Task 1}), it performs worse other unseen sequences in \textit{Task 2}. This is also evident in Fig.~\ref{fig:vis-result}(b), where SSLSAM performs better on sequences similar to its pretraining data but shows noticeably worse performance on others, such as SUB.
For MedSAM, we observe substantial performance drops across all tasks for breast and FGT. 
This likely results from its pretraining on a limited set of medical imaging datasets (approximately 1 million images, compared to 11 million natural images used for SAM), none of which included breast MRIs or CT, as indicated in the original MedSAM paper. However, it contains substantial data for abdominal MRIs, which makes it capable of capturing organ features effectively. As shown in Fig.~\ref{fig:feature}, its feature maps primarily emphasize organ structures.
MedCLIP-SAM fails across all tasks despite being trained on MRIs. Due to the low resolution of its feature maps (Fig.~\ref{fig:feature}), it is far from suitable for registration.

Overall, we do not observe clear benefits from using medical-domain pre-trained models for these breast MRI registration tasks. One possible explanation is that these models, despite being trained on medical data, still use much smaller datasets compared to large-scale natural image pre-training used by models like DINO-v2 or SAM, resulting in less generalizable features. 
We do not draw definitive conclusions at this stage regarding the effectiveness of medical pre-trained models for zero-shot registration tasks, as new models continue to be proposed \cite{sun2025foundation,dong2025mri}. Further exploration and discussion are needed to better understand and improve foundation models for this application.

\noindent\textbf{Discussion about ablation study.} 
We conducted two groups of ablation studies. The first explores the effect of PCA dimension ($d$) on registration performance, as shown in Fig.~\ref{fig:feature}. We observed that when $d$ is too low, it may discard important anatomical details necessary for accurate registration, as illustrated by the first three dimensions in Fig.~\ref{fig:feature} (c). However, as $d$ increases, the performance plateaus and reaches a relatively stable level. Since larger $d$ values significantly increase memory usage and computation time, we find that selecting $d$ in the range of 10–16 offers a good trade-off for each encoder.
The second study evaluates the contribution of feature maps and optimizers to registration performance. As shown in Table~\ref{tab:quant_result}, SAM’s extracted features outperform raw images, and the combination of SAM features with the ConvexAdam optimizer achieves the best performance among all VFM-based registration methods.

\noindent\textbf{Conclusion.} In this study, we explored whether foundation models can be used for zero-shot registration tasks in the context of breast MRI. We find that models pre-trained on natural images, such as SAM, can achieve strong performance for large-structure alignment, particularly for breast contour registration. However, these models still struggle to preserve fine anatomical details, such as FGT, limiting their application in tasks that require accurate tracking of internal structures. We believe this highlights an important direction for future research, which could develop strategies to better preserve fine-grained information within the feature representations of foundation models.

\noindent\textbf{Acknowledgments}
The research reported in this publication was supported by the National Institute of Biomedical Imaging and Bioengineering of the National Institutes of Health under the award number R01EB031575. The content is solely the responsibility of the authors and does not necessarily represent the official views of the National Institutes of Health.

\clearpage

\bibliographystyle{splncs04}
\bibliography{Paper-0047}

\end{document}